\documentclass[letterpaper, 10 pt, conference]{IEEEConf}
\IEEEoverridecommandlockouts

\usepackage{tikz}
\usepackage{cite}
\usepackage{amsmath,amssymb,amsfonts}
\usepackage{algorithmic}
\usepackage{graphicx}
\usepackage{textcomp}
\usepackage{xcolor}
\usepackage{listings}
\usepackage[frozencache,cachedir=.]{minted}
\usepackage{parcolumns}
\usepackage{hyperref}

\usepackage{todonotes}

\makeatletter
\let\@float@c@listing\@caption
\makeatother

\newcommand\copyrighttext{%
  \footnotesize \textcopyright 2024 IEEE.  Personal use of this material is permitted.  Permission from IEEE must be obtained for all other uses, in any current or future media, including reprinting/republishing this material for advertising or promotional purposes, creating new collective works, for resale or redistribution to servers or lists, or reuse of any copyrighted component of this work in other works.}
\newcommand\copyrightnotice{%
\begin{tikzpicture}[remember picture,overlay]
\node[anchor=south,yshift=10pt] at (current page.south) {\fbox{\parbox{\dimexpr\textwidth-\fboxsep-\fboxrule\relax}{\copyrighttext}}};
\end{tikzpicture}%
}

\def\BibTeX{{\rm B\kern-.05em{\sc i\kern-.025em b}\kern-.08em
    T\kern-.1667em\lower.7ex\hbox{E}\kern-.125emX}}

\title{\LARGE \bf
Enabling Waypoint Generation for Collaborative Robots \protect\\using LLMs and Mixed Reality}

\author{
Cathy Mengying Fang\textsuperscript{1,\textsection},
Krzysztof Zieliński\textsuperscript{2,3,\textsection},
Pattie Maes\textsuperscript{1},\\
Joe Paradiso\textsuperscript{1},
Bruce Blumberg\textsuperscript{3},
Mikkel Baun Kjærgaard\textsuperscript{2}
\thanks{\textsuperscript{1}Authors are with
MIT Media Lab
        {\tt\small catfang, pattie, joep@media.mit.edu }}%
\thanks{\textsuperscript{2}Authors are with
the Faculty of Engineering, the Maersk Mc Kinney Moller Institute, University of Southern Denmark
        {\tt\small krzi, mbkj@mmmi.sdu.dk }}%
\thanks{\textsuperscript{3}Authors are with Universal Robots A/S
        {\tt\small krzi, brbl@universal-robots.com}}%
\thanks{\textsuperscript{\textsection} Equal contribution}%
}

\begin{document}

\maketitle
\copyrightnotice
\thispagestyle{empty}
\pagestyle{empty}

\begin{abstract}
Programming a robot is a complex task, as it demands the user to have a good command of specific programming languages and awareness of the robot's physical constraints. We propose a framework that simplifies robot deployment by allowing direct communication using natural language. It uses large language models (LLM) for prompt processing, workspace understanding, and waypoint generation. It also employs Augmented Reality (AR) to provide visual feedback of the planned outcome. We showcase the effectiveness of our framework with a simple pick-and-place task, which we implement on a real robot. Moreover, we present an early concept of expressive robot behavior and skill generation that can be used to communicate with the user and learn new skills (e.g., object grasping).

\end{abstract}

\section{Introduction}

Manual programming has been the conventional way to set up industrial and collaborative robots – using a specific input controller to define various robot tasks, such as setting waypoints, connecting I/O ports to auxiliary equipment, creating program logic, etc. Usually, each robot arm manufacturer has a bespoke controller and programming language, which is often presented as a simple ladder diagram or graphical block representation. To overcome the limitations of robot-specific interface/language, more abstract programming methods have been developed, such as the Robot Operating System (ROS) which aims to provide a universal programming method.

The methods mentioned above require the robot programmer to have some basic computer science and robotics skills to grasp concepts such as singularities, joint space, inverse kinematics, etc. However, most of the world's manufacturing is done by Small and Medium Enterprises (SMEs) who often do not have such skills, which limits their ability to automate their processes.
To lower the barrier for robot programming, more automated methods exist~\cite{ajaykumar_survey_2022}, such as Programming by Demonstration (PbD) which allows users to demonstrate robot skills to the system, rather than coding them.

In addition to PbD, recent advances in generative AI, specifically the creation of foundation models, allow researchers to solve diverse robotics tasks not confined by the tasks, environments, and robot types in the prior training data \cite{kawaharazuka2024real}. This presents an immense opportunity not only for scaling robotic operations but also for improvements in human-robot interaction, given the semantic nature of the I/O of the large language models. Specifically, this could further lower the barrier of laypersons to operate a collaborative robot. 

In this paper, we propose a novel way to program a collaborative robot using natural language leveraging LLMs. Our framework takes in the speech input from the user and a 3D model of the physical scene and in real-time generates a series of waypoints. The generated waypoints can be simultaneously previewed in an augmented reality headset overlaid on top of the physical environment, and the robot receives the waypoints directly as a result of the generation and follows the approved plan. Furthermore, we also demonstrate an early exploration of automatic robotic skill generation, leveraging the generative nature of LLMs. We show that our framework can generate expressive gestures such as "nodding", and this capability can be generalized to other skills such as the "picking" skill in a pick-and-place task.

\begin{figure*}[t!]
    \centering
    \includegraphics[width=.9\textwidth]{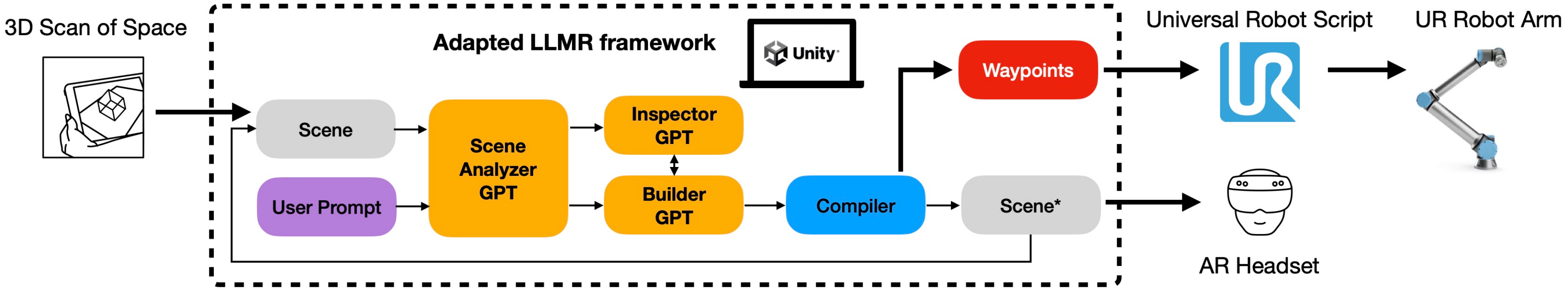}
    \caption{Overview of our framework. Starting from the top left, a 3D scene (optionally scanned, if it does not exist already) and a user prompt are fed to adapted LLMR framework, which is an orchestration of prompt-engineered GPT modules. The adapted framework outputs the trajectory based on the user prompt, which is converted to Universal Robots script, readable by the robot arm. The user also sees the rendered trajectory in the AR headset. }
    \label{fig:framework}
    \vspace{-15px}
\end{figure*}

\section{Related Work}


We start the Related Work section with current approaches to automating robot programming – "Programming by Demonstration". We then touch upon the use of Mixed Reality in assisting robot programming. Finally, we conclude with recent breakthroughs and experiments using foundation models like LLMs to scale up robotic tasks.

\subsection{Programming by Demonstration Interfaces}
There exist different ways to gather and transmit the trajectory data to the robotics system. For robot arms, it is typically kinesthetic teaching~\cite{saveriano_incremental_2015}, where the user holds the robot’s end-effector and moves it along the desired trajectory. Takano et al.~\cite{takano_real-time_2016} focus on learning and reproducing basic motion primitives. Caccavale et al.~\cite{caccavale_kinesthetic_2019} additionally introduce structured cooperative tasks from human demonstrations. Kim et al.~\cite{kim_dual-arm_2016} implement three layers of kinesthetic teaching (joint, task and contact) for dual-arm robot control.

Kinesthetic teaching requires the human user to use both hands to manipulate the robot arm, which becomes more troublesome when the robot has more limbs, e.g., a humanoid robot. For that case, motion capture based approaches can be used to imitate human body movements~\cite{kulic_incremental_2012,ude_programming_2004}. In this setup, the human teacher is equipped with sensors that are externally tracked with optical or magnetic tracking systems. The major challenge with this approach is the correspondence problem where the demonstrator and imitator do not have the same kinematic and dynamic properties (e.g., human vs. humanoid). Alternatively, prior work used different mappings ~\cite{skoglund_programming-by-demonstration_2010}  or tools ~\cite{hanifi_elhachemi_amar_trajectory_2020, maric_unsupervised_2020} to mediate the difference in the properties. 

Teleoperation does not face the correspondence problem (i.e., the operator is constrained to the robot's sensors), and it also allows the operator to perform the task remotely. The human teacher requires some type of input device and feedback such as joysticks ~\cite{yang_repeatable_2017} or custom control devices ~\cite{kuklinski_teleoperation_2014, kitagawa_online_2023} to operate the robot. 
\textit{Mobile ALOHA}~\cite{fu2024mobile} introduces whole-body teleoperation where the user is physically tethered to the device and can control both arms and the wheels.

Speech is another interface for PbD, although not so popular due to its ambiguities and limited command dictionary for robot control. Norberto~\cite{bugmann_robotbyvoice_2005} and Tasevski et al.~\cite{tasevski_integration_2013} use voice-controlled industrial robot to perform pick-and-place and welding tasks with a fixed set of voice commands.
We propose a novel way for automatic robot programming as it utilizes speech for robot control, and hence it does not force the user to memorize a constrained set of robot commands.

\subsection{Augmented Reality in Robotics}
Augmented Reality (AR) is a well-researched topic as an interface for Human-Robot Interaction~\cite{suzuki2022augmented}, robot control and teleoperation, etc. Quintero et al.~\cite{quintero_robot_2018} use a head-mounted display to allow users to create and preview robot trajectory. Gong et al.~\cite{gong_projection-based_2019} create robot instructions using projection-based AR for grasping tasks. While Stadler et al.~\cite{stadler_augmented_2016} explore handheld AR interface for industrial robot programming. Zielinski et al.~\cite{zielinski_study_2021} augment the robot arm with contextual data, such as software limits, joint temperatures, end-effector coordinate frame, etc. Chen et al.~\cite{chen20233d} control mobile robot with a drag-and-drop interface within a head-mounted AR display. Our work directly integrates speech and solution generation within the mixed-reality overlay, giving the user direct command of the robot.


\subsection{Situating LLM in the Real World for Robotics }
Kawaharazuka et al. overview the application of foundation Large Language Models (LLM) and Vision-Language Models (VLM) in real-world robotics~\cite{kawaharazuka2024real}. \textit{Palm-E} is one such example of a general-purpose multimodal language model that is trained for embodied reasoning (e.g., robot planning) \cite{driess2023palm}. Besides training new foundation models with new modalities, many can already solve many challenges in robotics by adapting existing foundation models. \textit{ProgPrompt} uses LLM to help with task planning and generate a correct sequence of steps in different contexts~\cite{singh2023progprompt}. Ahn et al. focus on a key limitation of using LLM to drive robotic behavior, which is the awareness of the robot's capabilities. They ground the output to the capabilities of the robot such that the output is actually feasible to execute~\cite{ahn2022can}. \textit{AutoRT} solves the challenge of lack of data grounded in the physical world by scaling up the deployment of auto-instructing robots using existing foundation models~\cite{ahn2024autort}. On the other hand, \textit{SpatialVLM} solves the same problem by developing an automatic 3D spatial visual QA data generation framework~\cite{chen2024spatialvlm}. Despite recent advances, the current models are still not suitable for tasks that require collaboration with humans~\cite{kawaharazuka2024real}.

Our work hopes to contribute to the improvement of human-robot collaboration and interaction. We do so by building on top of the Large Language Model for Mixed Reality (LLMR) framework~\cite{de2023llmr}, a framework that allows real-time natural language prompting of interactive 3D scenes. We take advantage of the ease of integration of the LLM-based framework within the mixed reality environment. The user can instruct the robot arm with natural language and preview the results directly in an AR headset, for example.

\begin{figure*}
    \centering
    \includegraphics[width=1\textwidth]{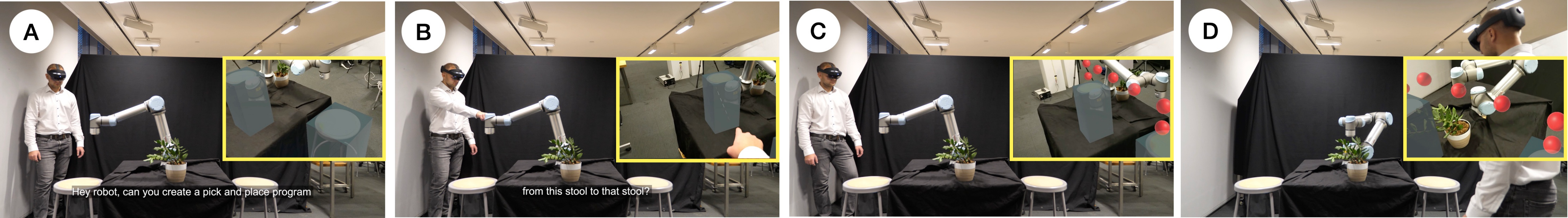}
    \caption{Example interaction between the user and the collaborative robot arm enabled by our framework. Mixed Reality views are outlined in yellow. A-B: The user is wearing a HoloLens2 AR headset and instructs the robot to create a pick-and-place program between two stools. C: Our framework (running within Unity on a separate laptop) generates a series of waypoints (indicated as the red spheres) and the waypoints are streamed to preview in the AR headset. D: Once the user is satisfied with the waypoints, the robot receives the command from Unity and then follows the waypoints.}
    \label{fig:interaction}
    \vspace{-15px}
\end{figure*}

\section{Framework overview}
Our framework encompasses the integration of LLMs, the Unity game engine, the HoloLens 2 AR headset, and URScript (proprietary Universal Robots' programming language) to achieve real-time AI-assisted planning of robot waypoints (\autoref{fig:framework}). We now discuss each component of the framework in detail.

\begin{figure}[b!]
    \centering
    \vspace{-15px}
    \includegraphics[width=.9\linewidth]{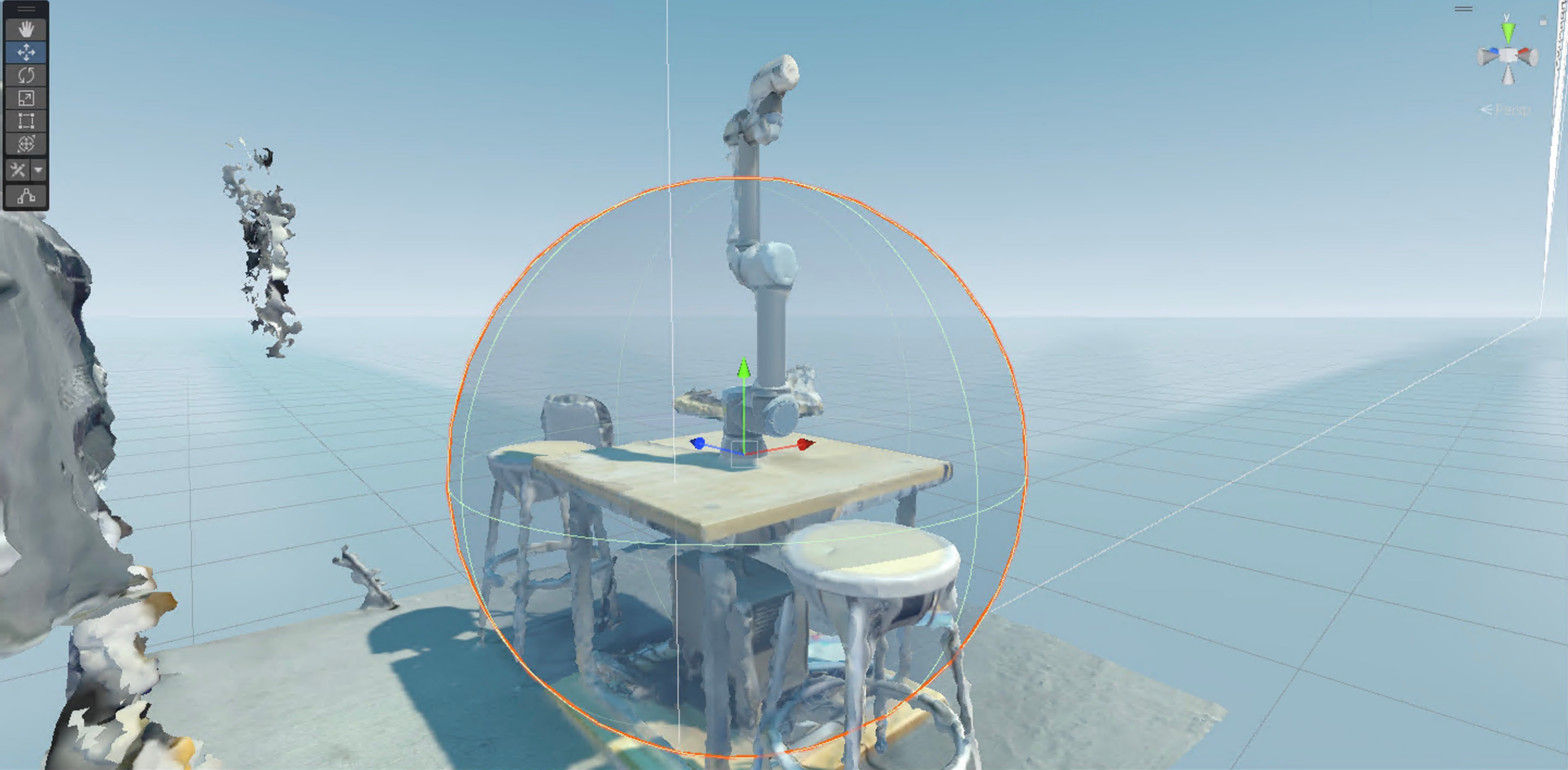}
    \caption{A pre-scanned scene and a model of the robot arm are loaded in the Unity environment. A translucent reachability sphere is shown to indicate the conservative estimation of the maximum reach of the robot arm.}
    \label{fig:unity}
\end{figure}

\textit{Digital representation of the physical scene} – We start by creating a replica of the physical scene in Unity, a rendering engine \cite{unity}. We used Polycam \cite{polycam} to capture a 3D model of the real-world objects and environment, and we imported the 3D model of the Universal Robots (UR10e) \cite{UR10} (\autoref{fig:unity}). The user can alternatively provide a scan of the physical world (most industrial plants likely have such models). The scan is used as a reference for the collider boundaries we manually configured in Unity, a manual step that can be replaced with a sophisticated scene understanding model \cite{sceneunderstanding}. We named the scene objects and the robot's sub-components according to their semantic meaning. We also set up a reachability sphere that represents the conservative upper bound of the robot endpoint's reach. We assume the physical objects are static (i.e., they do not move when the robot is moving and thus are not tracked). Future work can use real-time capture methods (e.g., OptiTrack or Aruco Markers) to update the scene in Unity. Combined with the user prompt, the scene is provided to our LLM system to generate a series of waypoints, which we discuss next.

\textit{LLM integration} – We built upon LLMR~\cite{de2023llmr}. Briefly here, LLMR is an orchestration of LLM modules prompt-engineered to achieve successful and flexible generations of interactive objects and scenes in Unity (Figure \ref{fig:framework}). LLMR is designed for simulating scenarios in virtual environments; we pushed the real-world implication of LLMR by connecting the simulated outcome to a physical robot. We fine-tuned a specific prompt for LLMR that provides the context of the task. An example prompt is in the Appendix \ref{appx:prompt}. We also optimized and improved the Scene Analyzer component in LLMR, designed to succinctly summarize the scene conditioned on the user request, to pay more attention to the robot and the relevant objects. The outputs of this step are the code that generates a series of sequential waypoints and the visualization of these waypoints in the virtual scene. A sample of the generated code is in the Appendix \ref{appx:waypoint_output}. In the iteration process of fine-tuning the prompt, we found that the generated trajectory pattern does not vary significantly across runs unless the prompt describes very specific constraints (such as specific obstacles to avoid or constraints of the robot arm motion).

\textit{Visualization of the scene in AR} – We deploy our framework's run-time compiler on a PC that acts as the server, and we build upon existing streaming protocols (holographic remoting for a HoloLens 2)  \cite{remoting, render_streaming} to stream the generated results to the head-mounted display.

\textit{Virtual robot control in Unity} – For the system to test if a point is reachable or not, we have implemented a simplified inverse kinematics (IK) controller in Unity. As an argument, it takes waypoint positions in Cartesian space w.r.t. (with reference to) the robot base and returns joint angles. The controller is based on gradient descent to calculate the trajectory between waypoints but does not account for singularities.

\textit{Real robot control from Unity} – To run our framework on the real robot, we used Universal Robots' proprietary Primary Interface for two-way communication with the arm. Once we obtain satisfactory waypoints, we can directly upload them to the robot. In this mode, there is no need to provide custom inverse kinematics, since we utilize Universal Robots' control box for trajectory planning in-between waypoints.

\begin{figure*}
    \centering
    \includegraphics[width=\textwidth]{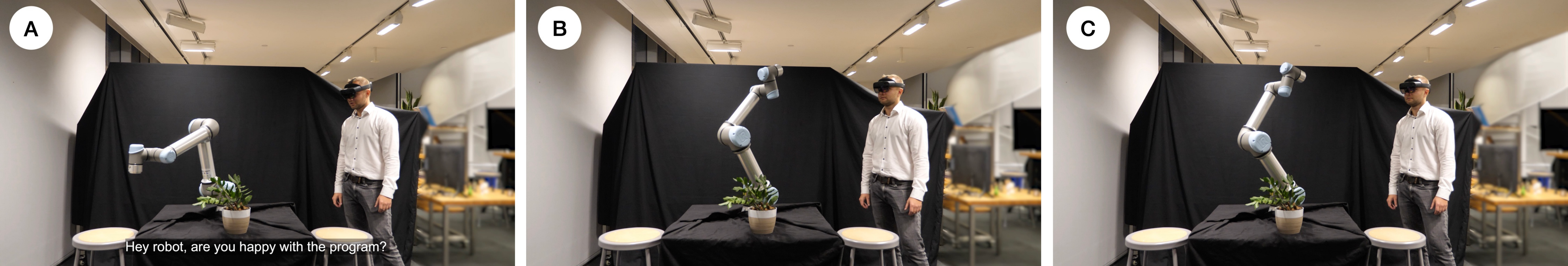}
    \caption{An example of an expressive response from the robot back to the user. A: The user asks the robot if it is happy with the generated program. B-C: The robot responds by nodding back at the user, where the nodding animation is generated by our framework.}
    \label{fig:expression}
    \vspace{-15px}
\end{figure*}

\section{Example Scenario}
A common task is pick-and-place, where the robot arm needs to move between two endpoints repeatedly while avoiding colliding with objects in the room or itself. Typically, the user would need to manually program the robot (first in simulation and then in the physical environment).

We envision the following end-to-end interaction where the user can achieve this with natural language instructions. The user wants to create a custom pick-and-place program between two places. They can point at the two locations and say "between this stool and that stool" (\autoref{fig:interaction} A-B). Our framework takes in the user prompt and the physical scene to create a detailed instruction. The instruction and scene information go through the LLM framework and the framework outputs a series of waypoints that are then visualized in the Augmented Reality headset (visualized as the red spheres, \autoref{fig:interaction} C). The user is able to see and verify the proposed waypoints in situ. After the user confirms, the waypoints are directly sent to the robot, where the robot's endpoint follows the generated sequence (\autoref{fig:interaction} D).

\section{Early Exploration: Few-shot Generation of Expressive Robot Behavior and Skills}

Overlaying visualization of the robot's trajectory in the real world is one of the ways that helps human users to directly manipulate and communicate with the robot. However, there lacks a channel and modality, such as gesturing \cite{mahadevan2024generative}, for the robot to communicate back to the human user.  We were also inspired by how the collaborative robot can be programmed to perform alongside a human dancer \cite{dancewithrobot}.

Following the extended animation generation capability of LLMR \cite{huang2023real}, we also experimented with using LLM to generate anthropomorphic expressions of the robot arm. We provided some context about the robot's joint limitations and a few examples of hand-crafted animations (Appendix \ref{appx:animation_prompt}). We used the gpt4-1106-preview model with a temperature of 0.1. A sample of the generated joint positions is in the Appendix \ref{appx:animation_output}. We imagine a scenario where the robot is asked whether they are pleased with the result, they could nod or shake their head back to the user (\autoref{fig:expression}).  We consider "nodding" and "shaking" the head to be similar to the skill of "grasping". Different from other LLM-based generative approaches that interpolates between different pre-existing skills, we generate new anthropomorphic animations with a few-shot example in the metaprompt. The resulting animations are still exploratory, as the robotic arm's inherent limitation in degrees of freedom cannot result in complex expressions. In addition, unlike animals or humans where ample examples of different behaviors exist, robot expressions are more idiosyncratic, and thus the quality of the generated output is up to the human's interpretation.

\section{Limitations and Future Work}

\textit{Singularities} – The current system is unaware of the singularities that might emerge from the generated trajectory. The reachability sphere is one way that we prevent the robot from going beyond its range of motion. We used the Universal Robots' IK to allow the robotic arm to reach between the endpoints. Our framework could be incorporated as part of a path-planning pipeline to eliminate singularities.

\textit{Trajectory Optimization} – Our system only generates one possible path during each run. When the LLM's temperature is set to zero, the generated solution is mostly consistent (as a higher temperature usually results in a non-deterministic behavior). Unless the physical constraints of the robot or the target positions change, the LLM's solution stays consistent across generations. However, we do not enforce any optimization parameters, such as ensuring the shortest cycle time or least energy consumption. Our solution is not meant to be compared with the mathematical approach to derive the most optimized path. Rather, focus on creating a draft trajectory that is directly in the place of the user's working environment. Future work might combine the advantages of both approaches.

\textit{Feedback from the human} – An immediate improvement to our framework is incorporating direct feedback from the human. Specifically, the human can not only directly edit the trajectory by dragging the generated waypoints in the Augmented Reality scene but also "teach" the robot why the changes are preferred. Akin to PbD, the user has the ultimate say in what the trajectory looks like, but our framework would allow the user to overview the entire trajectory in situ and use language to provide context. A user study is merited to understand the ideal way to engage the user in this process such that the agency of the user is preserved in AI-assisted path planning. 

\textit{Feedback from the world} – Our current system requires manual alignment between the physical and virtual scene, which can be easily solved using trackable markers \cite{hl2cv,qrcode}. However, if an object moves in the scene, the changes would need to be synchronized and updated in the virtual scene. The current computer vision based solution systems lack sophisticated tracking of 3D objects, and running a machine learning model for object classification is computationally expensive. One workaround is to use a combination of vision and language to query the scene \cite {gpt4v} or co-embed text and image pixels in the CLIP feature space to achieve scene understanding \cite {Peng2023OpenScene}. 

Another integration would be sensor data from external sensors deployed in the real world scene. At the moment, the robot arm stops when it senses that it has hit something, it would be helpful to incorporate these types of information in the LLM system and to further equip LLMs the abilities to perceive the world.

\section{Conclusion}

We introduce an end-to-end framework that simplifies the programming of collaborative robots for users without much prior knowledge. Building on top of the existing framework, LLMR, that can create interactive scenes in a simulated mixed-reality environment with natural language input, we adapted the input prompt such that a user can use natural language to instruct a collaborative robot. Our adapted framework supports the initial step of path planning, which is the generation of waypoints. The framework considers the unique constraints coming from the environment (e.g., avoid colliding with physical objects) and the robot itself (e.g., maximum reach). We also show initial exploration of the framework's ability to generate new "skills" in the form of animations with few-shot examples. The contribution of this work lies in its potential to leverage existing LLMs to work with embodied systems and environments at human-scale. We propose that this work bridges the gap between the human's and the robot's perception of the world with the ability for both to speak the same "language" (code) in the mixed reality environment.

\bibliographystyle{ieeetr}
\bibliography{ref}

\newpage
\appendices
\section{Prompt for the adapted LLMR framework}
\label{appx:prompt}
An example of a modified user prompt that allows the framework to understand the scene and instructions.

\begin{minted}[breaklines, breakanywhere, frame=single, breaksymbolleft=, fontsize=\footnotesize]{text}

You are controlling a robot arm named UR10e. Your goal is to find at least 5 waypoints that allows the robot end point to move from the TOP surface of stool 1 to the TOP surface of stool 2. The first and final waypoint should be the start and end destinations. The middle waypoints should allow the robot to travel in air but these waypoints should avoid colliding with all objects (like table) in the scene. The middle waypoints should form some sort of arc or curve. The robot's end point can not reach beyond the reachability sphere. You should first establish the mesh boundaries of the objects in the scene. Do not move the robot. Visualize the final waypoints as small red sphere with unique names.

\end{minted}

\section{Example code output of the LLM framework}
\label{appx:waypoint_output}

\begin{minted}[breaklines, breakanywhere, frame=single, fontsize=\footnotesize]{csharp}

using UnityEngine;

public class WaypointsGenerator : MonoBehaviour
{
    public GameObject stool1;
    public GameObject stool2;
    public GameObject table;
    public GameObject reachabilitySphere;
    public GameObject robotEndPoint;
    public GameObject waypointPrefab;
    public int waypointCount = 5;
    private Bounds stool1Bounds;
    private Bounds stool2Bounds;
    private Bounds tableBounds;
    private Bounds reachabilitySphereBounds;

    void Start()
    {
        summary = "This script generates waypoints for the robot end point to move from the top of stool 1 to the top of stool 2, avoiding collision with the table.";

        stool1 = GameObject.Find("Stool_1");
        stool2 = GameObject.Find("Stool_2");
        table = GameObject.Find("Table");
        reachabilitySphere = GameObject.Find("ReachabilitySphere");
        robotEndPoint = GameObject.Find("Robot_EndPoint");

        waypointPrefab = GameObject.CreatePrimitive(PrimitiveType.Sphere);
        waypointPrefab.name = "WaypointPrefab_" + System.Guid.NewGuid().ToString();
        waypointPrefab.GetComponent<Renderer>().material.color = Color.red;
        waypointPrefab.transform.localScale *= .1f;
        Destroy(waypointPrefab); // We only needed it for the prefab

        stool1Bounds = GetMeshSize.GetRenderBounds(stool1);
        stool2Bounds = GetMeshSize.GetRenderBounds(stool2);
        tableBounds = GetMeshSize.GetRenderBounds(table);
        reachabilitySphereBounds = GetMeshSize.GetRenderBounds(reachabilitySphere);

        GenerateWaypoints();
    }

    void GenerateWaypoints()
    {
        Vector3 start = stool1Bounds.center + new Vector3(0, stool1Bounds.extents.y, 0);
        Vector3 end = stool2Bounds.center + new Vector3(0, stool2Bounds.extents.y, 0);

        for (int i = 0; i < waypointCount; i++)
        {
            float t = (float)i / (waypointCount - 1);
            Vector3 waypointPosition = Vector3.Lerp(start, end, t);

            // Make sure the waypoint is above the table and within the reachability sphere
            if (waypointPosition.y <= tableBounds.max.y || !reachabilitySphereBounds.Contains(waypointPosition))
            {
                waypointPosition.y = Mathf.Max(waypointPosition.y, tableBounds.max.y + 0.1f);
                waypointPosition = reachabilitySphereBounds.ClosestPoint(waypointPosition);
            }

            // Create the waypoint
            GameObject waypoint = Instantiate(waypointPrefab, waypointPosition, Quaternion.identity);
		Debug.Log(waypointPosition);
            waypoint.name = "Waypoint_" + i;
        }
    }
}

\end{minted}

\section{animation prompt}
\label{appx:animation_prompt}

\begin{minted}[breaklines, breakanywhere, frame=single, breaksymbolleft=, fontsize=\footnotesize]{text}

You're an animator who will be provided the joints on a rigged 3D model of a robotic arm, and you have to rotate the joints to produce the requested animation. The joints will be given as a JSON string that outlines the object hierarchy. 
You have to output a string that specifies the rotation of movement for each joint.

Each subsequent line of the string starts with the joint name, then each vector in the format (timestamp,rotation), specifing a key frame for the animation. Rotation is in radians.


# Example:
The object you will animate is a **robot**.
Object JSON:
name:Robot,children:[name:z-up,children:[name:root,children:[name:__base,rotation:(0.0),children:[name:__shoulder,rotation:(0.0),children:[name:__elbow,rotation:(0.0),children:[name:__wrist-1,rotation:(0.0),children:[name:__wrist-2,rotation:(0.0),children:[name:__wrist-3,rotation:(0.0)]]]]]]]]
__base clockwise direction: (1.0)
__shoulder clockwise direction: (1.0)
__elbow clockwise direction: (-1.0)
__wrist-1 clockwise direction: (1.0)
__wrist-2 clockwise direction: (1.0)
__wrist-3 clockwise direction: (1.0)

Instruction: animate the robot to bow

Robot/z-up/root/__base,(0.000,-1.571),(1.020,-1.571),(2.020,-1.571),(3.020,-1.571),(4.020,-1.571),(5.020,-1.571)
Robot/z-up/root/__base/__shoulder,(0.000,-1.571),(1.020,-1.571),(2.020,-1.570),(3.020,-1.570),(4.020,-1.570),(5.020,-1.571)
Robot/z-up/root/__base/__shoulder/__elbow,(0.000,0.000),(1.020,-0.472),(2.020,-1.519),(3.020,-2.089),(4.020,-1.358),(5.020,-0.315)
Robot/z-up/root/__base/__shoulder/__elbow/__wrist-1,(0.000,-1.571),(1.020,-1.239),(2.020,-0.504),(3.020,-0.104),(4.020,-0.617),(5.020,-1.349)
Robot/z-up/root/__base/__shoulder/__elbow/__wrist-1/__wrist-2,(0.000,1.571),(1.020,1.571),(2.020,1.571),(3.020,1.570),(4.020,1.571),(5.020,1.571)
Robot/z-up/root/__base/__shoulder/__elbow/__wrist-1/__wrist-2/__wrist3,(0.000,0.000),(1.020,0.000),(2.020,0.000),(3.020,0.000),(4.020,0.000),(5.020,0.000)


# Example:
The object you will animate is a **robot**.
Object JSON:
name:Robot,children:[name:z-up,children:[name:root,children:[name:__base,rotation:(0.0),children:[name:__shoulder,rotation:(0.0),children:[name:__elbow,rotation:(0.0),children:[name:__wrist-1,rotation:(0.0),children:[name:__wrist-2,rotation:(0.0),children:[name:__wrist-3,rotation:(0.0)]]]]]]]]
__base clockwise direction: (1.0)
__shoulder clockwise direction: (1.0)
__elbow clockwise direction: (-1.0)
__wrist-1 clockwise direction: (1.0)
__wrist-2 clockwise direction: (1.0)
__wrist-3 clockwise direction: (1.0)

Instruction: animate the robot shaking as if it's saying no

Robot/z-up/root/__base,(0.000,-1.514),(1.000,-1.514),(2.000,-1.514)
Robot/z-up/root/__base/__shoulder,(0.000,-0.931),(1.000,-0.932),(2.000,-0.932)
Robot/z-up/root/__base/__shoulder/__elbow,(0.000,-1.281),(1.000,-1.281),(2.000,-1.281)
Robot/z-up/root/__base/__shoulder/__elbow/__wrist-1,(0.000,-0.904),(1.000,-0.904),(2.000,-0.904)
Robot/z-up/root/__base/__shoulder/__elbow/__wrist-1/__wrist-2,(0.000,1.172),(1.000,1.612),(2.000,1.752)
Robot/z-up/root/__base/__shoulder/__elbow/__wrist-1/__wrist-2/__wrist3,(0.000,0.000),(1.000,0.000),(2.000,0.000)

\end{minted}

\section{animation output}
\label{appx:animation_output}

\begin{minted}[breaklines, breakanywhere, frame=single, breaksymbolleft=, fontsize=\footnotesize]{text}

animate the robot saying yes
[0,0,0.5,0,0,0]
[0,0,1.0,0,0,0]
[0,0,0.5,0,0,0]
[0,0,0,0,0,0]
[0,0,0.5,0,0,0]
[0,0,1.0,0,0,0]
[0,0,0.5,0,0,0]
[0,0,0,0,0,0]

animate the robot purring like a cat
[0,0,0,0,0,1]
[0,0,0,0,0,0.5]
[0,0,0,0,0,1]
[0,0,0,0,0,0.5]
[0,0,0,0,0,1]
[0,0,0,0,0,0.5]
[0,0,0,0,0,1]
[0,0,0,0,0,0.5]


the robot is laughing hysterically
[0,0,0,0,0,0]
[0.2,-0.1,0.3,0.1,-0.2,0.3]
[-0.2,0.1,-0.3,-0.1,0.2,-0.3]
[0.2,-0.1,0.3,0.1,-0.2,0.3]
[-0.2,0.1,-0.3,-0.1,0.2,-0.3]
[0.2,-0.1,0.3,0.1,-0.2,0.3]
[-0.2,0.1,-0.3,-0.1,0.2,-0.3]
[0,0,0,0,0,0]


the robot is so disappointed and embarrassed
[0,0,0,0,0,0]
[0.2,-0.5,1.2,-0.8,0,0]
[0,-0.8,1.5,-1.2,0,0]
[0.2,-0.5,1.2,-0.8,0,0]
[0,-0.8,1.5,-1.2,0,0]
[0,0,0,0,0,0]
[0,-0.3,0.5,-0.3,0,0]
[0,0,0,0,0,0]
[0,-0.3,0.5,-0.3,0,0]
[0,0,0,0,0,0]

\end{minted}

\end{document}